\documentstyle[aps, prb, twocolumn, floats, epsfig]{revtex}

\begin{document} 
%
\preprint{NSF-ITP-00-121}
\title{Screening of long-range Coulomb interactions in
the quasi two-dimensional extended Hubbard model:
A combined quantum Monte Carlo and Feynman diagram study}
\author{H.-B. Sch\"{u}ttler$^{1,2}$, C. Gr\"ober$^3$, H. G. Evertz$^{3,4}$,
        and W. Hanke$^{2,3}$\vskip2mm}
\address{$^1$ Center for Simulational Physics, Department of Physics
 and Astronomy, University of Georgia, Athens, Georgia 30602 }
\address{$^2$ Institute for Theoretical Physics
University of California,
Santa Barbara, California 93106, USA}
\address{$^3$ Institut f\"ur Theoretische Physik, Am Hubland,
Universit\"at W\"urzburg, D-97074 W\"urzburg, Germany }
\address{$^4$ Institut f\"ur Theoretische Physik,
Technische Universit\"at Graz, 8010 Graz, Austria}
\date{April 17, 2001}
\twocolumn[\hsize\textwidth\columnwidth\hsize\csname @twocolumnfalse\endcsname
\maketitle

\begin{abstract}
By combining fermion quantum Monte Carlo (QMC) simulations
with diagrammatic theory, we have calculated the
dielectric screening and the screened electron-electron 
interaction potential in a quasi two-dimensional 
extended single-band Hubbard model for cuprate superconductors,
with and without a long-range ($1/r$) extended Coulomb interaction potential,
at and near band-filling $1\over2$. Our results for the 
insulating dielectric screening in the $1\over2$-filled 
band case indicate that the Hubbard conduction
electron system contributes only a minor fraction, 
$\Delta\epsilon\cong 0.9$, of the total observed in-plane electronic
dielectric constant of the cuprates, $\epsilon_{\infty\parallel}{}\cong 4.7$.
With increasing hole doping concentration $x$, the extended ($1/r$-) part
of the Coulomb interaction potential is rapidly suppressed
by the metallic screening, due to doping induced charge carriers.
Surprisingly, near $x\sim5\%$, the low-frequency component
of the screened potential $V_S$ changes sign 
and becomes attractive at 1st neighbor and more extended distances
in the $CuO_2$-plane. The 1st neighbor screened potential
reaches maximum attraction strength at dopings $x\sim 13-15\%$ and 
and becomes repulsive again at larger dopings, for $x\sim 23-25\%$.
Similar results are found for the screened potential
of the pure 2D Hubbard model, without extended Coulomb 
interaction terms, suggesting that the extended Coulomb terms
do not qualitatively alter the local screening physics of the 
doped Hubbard electron system. When included exactly in the
screening calculation, the 1st neighbor extended Coulomb term 
enhances the on-site and 1st neighbor overscreening 
attraction already present in the pure Hubbard model at finite doping.
Our results are potentially relevant for the question of
the $d$-wave pairing mechanism in the cuprates,
since they suggest that the screened extended Coulomb interaction 
potential does not only not suppress a $d$-wave 
pairing state but, due to the overscreening, could actually 
increase the $d$-wave attraction, in the $5-25\%$ doping regime.
Our results may also have implications for the explanation 
of the isotope effect and its doping dependence in the cuprates.
At larger dopings, $x>15\%$, the screened 
potential becomes attractive even on-site, suggesting the possibility
that the screened potential could support 
or enhance $s$-wave pairing in the high-doping regime. 
We also give a rigorous analytical proof that the screened 
on-site interaction must become attractive
near half-filling in the repulsive large-$U$-limit
of both the pure Hubbard and the extended Hubbard single-band model.
We present a simple physical interpretation of this
seemingly paradoxical result in terms of the retardation
effects which are inherent in the screening of an instantaneous
repulsive interaction potential. We also point out that
on-site overscreening implies singularities in the
imaginary-frequency dependence of the irreducible 
polarization insertion and its 3-point vertex function.
\end{abstract}
]

\section{Introduction}
\label{sec:intro}
In the cuprate high-$T_{\rm c}$ superconductors, 
the strong, on-site Hubbard-$U$ Coulomb 
repulsion is believed to prevent
conventional on-site Cooper pair formation.
This long-standing theoretical dictum is
now supported by experimental evidence, 
obtained in several cuprate materials, for a non-$s$-wave
pairing state, of $d_{x^2-y^2}$-symmetry.\cite{vanharlingen,scalapino,new_ncco}
A $d$-wave pairing state implies
a spatially extended Cooper pair wavefunction,
involving electrons paired at 1st neighbor or larger 
lattice distances. 
Such a non-$s$-wave state
is impervious to the on-site $U$-repulsion,
but it can still be suppressed by the extended part
(1st, 2nd, ... neighbor) of the electron-electron 
Coulomb repulsion. The bare strength of this
extended Coulomb potential may be quite substantial
in the cuprates.
For any proposed microscopic model of the cuprates,
it is therefore crucial to demonstrate that
the model can support a $d$-wave pairing state
near $1\over 2$-bandfilling
and that such a pairing state is robust against 
extended Coulomb repulsions.

In the present paper, we combine
diagrammatic and quantum Monte Carlo (QMC)
methods to study both the insulating dielectric
screening at band-filling $1\over2$ and, as a function
of doping, the screened
electron-electron interaction potential 
near the $1\over2$-filled band limit within the
framework of a quasi two-dimensional (2D) Hubbard
model with three-dimensional (3D) extended
$1/r$ Coulomb interactions.
In the $1\over2$-filled band case, our results 
suggest that the insulating Hubbard
electron system contributes only a minor fraction, 
$\Delta\epsilon\cong 0.9$, to the total observed electronic
dielectric constant of the cuprates, $\epsilon_\infty{}\cong 4.7$,
with most of the electronic dielectric screening being
provided by non-Hubbard ``background'' electrons which
are not explicitly included in the conventional single-band
Hubbard description of the cuprates.
In comparing to the observed values for the
dielectric constant, we can thus obtain an estimate
for the strength of the $1/|r|$ Coulomb interaction potential
of the quasi 2D Hubbard electron system. 
At finite doping, we find that screening due 
to the doping induced "metallic" charge fluctuations 
causes the screened electron-electron interaction
potential $V_{\rm S}$ to become attractive.
At low doping concentrations $x$, 
this ``overscreening'' effect reverses the sign 
of the spatially extended part of $V_{\rm S}$, giving rise
to a 1st neighbor attraction in the $5-25\%$ doping
range, with a attraction strength maximum near $x=15\%$.
At larger $x$, even the on-site part of $V_{\rm S}$ 
turns attractive, for $x\gtrsim15\%$.
We give a rigorous analytical proof for the existence of this
on-site overscreening effect and show that it is intrinsically a doping induced
large-$U$ effect of Hubbard systems near band-filling $1\over2$.
Possible implications of these overscreening effects
for superconducting pairing instabilities in the 2D Hubbard
and quasi-2D extended Hubbard model are also discussed.

The remainder of the paper is organized as follows:
In Section~\ref{sec:model}, we describe our model and outline the basic 
approach of combining QMC simulations with diagrammatic
theory. In Section~\ref{sec:dielectric}, we present our calculation of the dielectric
constant in the $1\over2$-filled Hubbard electron system
and estimate the ``bare'' strength of the extended $1/|r|$ Coulomb
interaction, by comparing to experimentally observed values
in insulating cuprate parent compounds. In Section~\ref{sec:v_screened}, we
present our results for the on-site and 1st neighbor
component of the screened Coulomb potential in the 2D Hubbard
and quasi-2D extended Hubbard model. In Section~\ref{sec:ex_proof}, we present
an exact proof for the existence of on-site overscreening
in the asymptotic $U\to\infty$ limit and discuss its implications
for the analytical structure of the irreducible polarization insertion
and its underlying vertex function. In Section~\ref{sec:pairing}, we speculate
on possible implications of overscreening for the superconducting
pairing instabilities of the 2D Hubbard and quasi-2D extended Hubbard model.
Section~\ref{sec:summary} contains a brief summary.
%
%
%
\section{Model and combined diagrammatic quantum Monte Carlo approach}
\label{sec:model}
Both theoretical (bandstructure) calculations
and experimental observations, specifically the highly 
anisotropic electronic transport properties of the cuprates,
suggest that their electronic stucture is close to two-dimensiomnal
and it is a reasonable first approximation to neglect interlayer 
electronic hybridization effects.
However, the dielectric screening in the undoped
insulating parent compounds is far less anisotropic, with
observed dielectric constants of quite similar magnitude
parallel and perpendicular to the $CuO_2$-layer directions.
This suggests that the extended $1/r$ part of the electronic
Coulonb repulsion is by no means confined to a single $CuO_2$-layer
and that the full $3D$ interaction potential may have to
be taken into account.
We therefore start from an extended Hubbard Hamiltonian of the form \cite{sf}
%
\begin{equation}
H\ =\ \sum_{j,\ell} 
    \Big(
   {1\over2} V(r_{j\ell}) n_j n_\ell
   -\small\sum_{\sigma} t_{j\ell}
    c_{j\sigma}^\dagger c_{\ell\sigma}
    \Big) 
%
 \ \equiv\ H_V + H_t
\;,
\;\label{eq:model}
\end{equation}
%
with $c_{j,\sigma}^\dagger$ creating an 
electron of spin $\sigma=\uparrow,\downarrow$
at $Cu$-site $r_j$ in a three-dimensional (3D) crystal 
of stacked $CuO_2$ layers,
$n_j=\sum_\sigma c_{j\sigma}^\dagger c_{j\sigma}$ and 
$r_{j\ell}=r_j-r_\ell$. $H_t$ includes only an in-plane
1st neighbor hybridization $t$ and the chemical
potential $\mu$. Some preliminary results including
a 2nd neighbor hybridization $t^{\bf\prime}$ have also been obtained.
The 3D Coulomb potential
\begin{equation}
V(r)\ =\ U\delta_{r,0} +{e^2\over \epsilon_B |r|_{\rm min}} (1-\delta_{r,0})
\ \equiv\ U\delta_{r,0} + V_{\rm e}(r)
\label{eq:coulomb}
\end{equation}
includes the on-site repulsion $U$ and an 
extended $1/|r|$-part, $V_{\rm e}$, with a dielectric constant 
$\epsilon_B$ to account for screening by
the insulating background not explicitly included in $H$,
that is "non-Hubbard" electrons in lower
filled bands and, possibly, phonon degrees of freedom.
On a finite $L^{(1)}\times L^{(2)}\times L^{(3)}$ 
lattice with periodic boundary conditions and
primitive unit vectors $r^{(\alpha)}$, having linear
dimension of $L^{(\alpha)}$ unit cells in the $r^{(\alpha)}$-direction 
for $\alpha=1,2,3$, we define
\begin{equation}
|r|_{\rm min} = \min_{m_1,m_2,m_3} |r-\sum_{\alpha=1}^3 m_\alpha L^{(\alpha)} r^{(\alpha)}|
\end{equation}
with the $m_\alpha$ taken over all integers, to ensure proper periodicity
of $V(r)$.

The exact screened potential $V_{\rm S}$, the exact irreducible polarization 
insertion $P$ and the exact density correlation 
function $\chi$ of the full Hamiltonian (\ref{eq:model}) 
are related by\cite{mahan}
\begin{eqnarray}
V_{\rm S}(q,i\omega) 
&=& [1-V(q) P(q,i\omega)]^{-1} V(q)
\nonumber\\
&=& V(q) - V(q)\chi(q,i\omega) V(q)
\;,\label{eq:v_s}
\end{eqnarray}
where $V(q)$ denotes the lattice Fourier sum over $V(r)$ and 
\begin{eqnarray}
\chi(q,i\omega)&=& 
{1\over N} \sum_{j,\ell}\int_0^\beta d\tau
e^{i\omega\tau-iq\cdot r_{j\ell}}
\langle \Delta n_j(\tau) \Delta n_\ell(0) \rangle  
\nonumber\\
%
&\equiv& -P(q,i\omega)[1-V(q)P(q,i\omega)]^{-1}
\label{eq:chi}
\end{eqnarray}
at wavevectors $q$, Matsubara frequencies $i\omega$
and temperature $T\equiv 1/\beta$ for lattice size $N$
with $\Delta n_j\equiv n_j-\langle n_j\rangle$.\cite{mahan}
Here, $\langle\dots\rangle$ and $\dots(\tau)$ denote, respectively,
thermal averaging and imaginary-time evolution with respect to $H$.
Note that $V_{\rm S}$ and $P$ depend on the choice of
representation for $H_V$. Eqs.~(\ref{eq:v_s},\ref{eq:chi})
are based on the diagrammatic expansion in the {\it charge} representation
where the $U$-term is written as ${U\over2}\sum_j n_j^2$,
rather than the more familiar {\it spin} representation 
$U\sum_j n_{j\uparrow}n_{j\downarrow}$. While both
are equivalent when all diagrams are summed exactly to all orders, 
the former, as we will discuss,
may offer some advantages for approximate diagram resummations.
 
The basic idea of our QMC approach is to
calculate $V_{\rm S}$, via Eq.~(\ref{eq:v_s}), from 
the polarization insertion $P$ which, in turn, 
is extracted, via Eq.~(\ref{eq:chi}), 
from QMC results for the density
correlation function $\chi$.
In order to include the full 3D extended
Coulomb repulsion into this
approach, we treat only a certain short-range
portion of $V(r)$, denoted by $V_o(r)$, 
exactly, by QMC methods.
The remaining weaker, but long-range
part of $V$, denoted by $V_{\rm f}(r)\equiv V(r)-V_o(r)$, 
is then handled perturbatively.

By retaining only in-plane terms in $V_o$, 
the QMC simulation can be restricted to a single 2D layer.
We thus calculate, by QMC, the density correlation function 
\begin{eqnarray}
\chi_o(q,i\omega) &=&
{1\over N} \sum_{j,\ell}\int_0^\beta d\tau
e^{i\omega\tau-iq\cdot r_{j\ell}}
\langle \Delta n_j(\tau)_o \Delta n_\ell(0)_o \rangle_o  
\nonumber\\
&\equiv& -P_o(q,i\omega)[1-V_o(q)P_o(q,i\omega)]^{-1}
\label{eq:chi_o}
\end{eqnarray}
for the QMC Hamiltonian 
\begin{equation}
H_o\equiv H_{V_o} + H_t
\label{eq:model_o}
\end{equation}
on a single 2D $CuO_2$-plane, with the interaction potential 
$V$ in Eq.~(\ref{eq:model}) replaced by $V_o$. 
Here, $\langle\dots\rangle_o$ and $\dots(\tau)_o$ denote, respectively,
thermal averaging and imaginary-time evolution with respect to $H_o$
and $P_o$ is the exact irreducible polarization 
insertion of $H_o$ \cite{mahan} which can be extracted from
the QMC results for $\chi_o$ via Eq.~(\ref{eq:chi_o}). 

Our essential approximation is to replace the exact $P$ 
of the full Hamiltonian $H$ in Eqs.~(\ref{eq:v_s},\ref{eq:chi}) by $P_o$,
{\it i.e.}, we set
\begin{equation}
P(q,i\omega) \cong P_o(q,i\omega) = [V_o(q) - 1/\chi_o(q,i\omega)]^{-1}
\label{eq:p_o}
\end{equation}
in Eqs.~(\ref{eq:v_s}) and (\ref{eq:chi}).
All renormalizations of $P$ due to the short-range part of the interaction
potential, $V_o$, are thus included exactly, to all orders of $V_o$. 
Renormalizations due to the weaker long-range part $V_{\rm f}$ are neglected in $P$,
but approximately included in $V_{\rm S}$, via Eq.~(\ref{eq:v_s}), and in $\chi$,
via Eq.~(\ref{eq:chi}), {\it i. e.} by setting
$V_{\rm S}(q,i\omega)\cong [1-V(q) P_o(q,i\omega)]^{-1} V(q)$ 
and
$\chi(q,i\omega)\cong -P_o(q,i\omega)[1-V(q) P_o(q,i\omega)]^{-1}$.
Note that this approximation for $P$ becomes exact if we replace $H$ by $H_o$,
as, {\it e.g.}, in our calculations of $P$ and $V_{\rm S}$ for the pure 
2D Hubbard model discussed below. 

As a simple cuprate system, we consider $La_{2-x}Sr_xCuO_4$,
with a body-centered tetragonal ($bct$) 3D model crystal structure,
in-plane lattice constant $a=3.80\AA$, inter-layer spacing
$d=6.62\AA$ \cite{jorgensen}, $t=0.35eV$ and $U=8t$ \cite{sf}. 
(Small orthorhombic or larger-unit-cell tetragonal distortions
from this idealized structure are neglected.) Note that the foregoing
values for $U$ and $t$ are consistent with the values derived
by the systematic mapping of the low-energy Hilbertspace of
the three-band onto the single-band Hubbard model.\cite{sf}
The values of $U$ and $t$ are also consistent with
(and, essentially, determined by) the observed 
values of the antiferromagnetic 
exchange constant $J\cong0.13{\rm eV}$ 
and of the Mott-Hubbard charge excitation 
gap $\Delta_{MH}\cong1.8-2.6{\rm eV}$
in the undoped $La_2CuO_4$ parent compound.\cite{sf}

Using standard finite-temperature fermion determinant QMC methods,
$\chi_o$ is simulated with up to $2\times10^7$ MC sweeps and typically
$\lesssim 0.5\%$ statistical error on $6\times6$, $8\times 8$
and $10\times 10$ 2D square lattices
with periodic boundary conditions at
$\beta\equiv1/T$ up to $\beta t=10$, 
with imaginary-time step 
$\Delta\tau\equiv\beta/L_\tau\!\leq\!0.0625t^{-1}$
where $L_\tau$ is the Trotter number.
In most simulations, we use for $H_o$ the pure Hubbard model with 
\begin{equation}
V_o(r)=U\delta_{r,0}
\label{eq:v_o0}
\end{equation}
In order to explore the effects arising from inclusion of extended $V$-terms
in the polarization insertion $P_o$, we have also performed a few simulations
for the 1st neighbor extended Hubbard model, with 
\begin{equation}
V_o(r) =U\delta_{r,0} + V_1\sum_{s\in{\cal N}_1}\delta_{r,s}
\label{eq:v_o1}
\end{equation}
where 
\begin{equation}
V_1\equiv {e^2\over \epsilon_B a}
\label{eq:v_1}
\end{equation}
is the strength of the bare in-plane 1st neighbor Coulomb repulsion
and ${\cal N}_1$ denotes the set of in-plane 1st neighbor lattice vectors,
$s=(\pm a,0,0)$ and $(0,\pm a,0)$.

\section{Dielectric screening in the $1\over2$-filled Mott-Hubbard insulator}
\label{sec:dielectric}
Our first objective is to calculate the dielectric 
constant of the insulator from our model and thereby 
obtain an estimate for the background screening $\epsilon_B$ 
[in Eq.~(\ref{eq:coulomb})] {}from the measured 
long-wavelength external field dielectric tensor 
$\widetilde{\epsilon}_{\rm ex}(\omega)|_{q\to0}$ 
in the undoped, insulating $La_2CuO_4$ material.\cite{eps_expa,eps_expb}
Note here that the observed 
$\widetilde{\epsilon}_{\rm ex}(\omega)|_{q\to0}$ 
contains contributions
from both the conduction electron system, explicitly included
in our Hubbard model description, {\it and} from the
``background'' electron and phonon degrees of freedom
which are not explicitly included in our single-band Hubbard description.
Note also that in our simplified model, $\epsilon_B$ is represented as a
momentum- and frequency-independent constant, {\it i.e.}, in effect
the background is treated as a structureless homogeneous
dielectric medium, down to atomic length scales. We should 
caution from the outset that such an approach 
can provide only reasonable order of magnitude estimates. 

In a more quantitative modeling approach, based {\it e.g.},
on a shell model representation of the 
insulating non-Hubbard electron and phonon
background,\cite{hbs-shell} one may include more realistically the
momentum and frequency structure of $\epsilon_B$. However,
as discussed briefly below, the ``local field'' effects
in such a model, arising from the momentum dependence of 
$\epsilon_B$, will not qualitatively alter our main conclusions.
Also, the frequency dependence of the 
electronic contribution to $\epsilon_B$ is likely
of lesser importance, due to the high charge excitation energy
scale of the background electrons which is expected\cite{sf}
to be in the several eV range, 
exceeding the Hubbard electron bandwidth $8t$.
Via the phonon contribution to $\epsilon_B$, 
on the other hand, the low phonon frequency scale 
$\Omega_{\rm ph}\lesssim 0.1{\rm eV}\ll 8t$,
enters into the screening of the Coulomb interaction,
and, along with it, isotopic mass dependence.
This may have implications for the explanation of the isotope
effect observed in the superconducting properties of the
cuprates, especially in the underdoped regime, as 
discussed further in Section~\ref{sec:v_screened}.
However, as we will also explain in Section~\ref{sec:v_screened},
neither the presence of the phonon contribution in $\epsilon_B$
nor its low frequency scale, will substantially alter the Hubbard 
electrons' ``metallic'' screening at finite doping
which is the central focus of the present paper. 

Within the framework of our simplified background dielectric model, 
we obtain the longitudinal (scalar) dielectric function 
$\epsilon_{\rm ex}$ from \cite{eps_th}
\begin{equation}
{1\over\epsilon_{\rm ex}(q,i\omega)} 
= {1\over\epsilon_B}
\Big[
1 - {4 \pi e^2\over \epsilon_B {\cal V}_{\rm c} |q|^2} \chi(q,i\omega)
\Big]
\label{eq:e_ex-q}
\end{equation}
at $i\omega=0$. Here, ${\cal V}_{\rm c}$ denotes the 3D unit cell volume and
as discussed above, $\chi=-P[1-V(q)P]^{-1}$ from (\ref{eq:chi}) with $P\!\cong\!P_o$. 
The components of the static dielectric tensor $\widetilde{\epsilon}_{\rm ex}$
in the long-wavelength limit can then be extracted from
\begin{equation}
\hat{e}\cdot \widetilde{\epsilon}_{\rm ex} \cdot \hat{e}
= \lim_{|q|\to 0}  \lim_{L\to\infty} \epsilon_{\rm ex}(|q|\hat{e},i\omega=0)
\label{eq:e_ex-tensor}
\end{equation}
for arbitrary unit vectors $\hat{e}$ with 
$L$ denoting the linear lattice size.
Note here that it is crucial to take the thermodynamic 
limit $L\to\infty$ at finite $|q|>0$ {\it before} 
taking the long-wavelength limit $|q|\to0$. 

Since the finite-system QMC simulations provide us 
with density correlation data $\chi_o(q,i\omega)$ 
only on a finite, discrete $q$-grid, 
some special care must be taken to extract the
foregoing limits from the QMC data. To do so, we have developed
a simple, physically motivated $r$-space embedding procedure
which provides us with a continuous $q$-space interpolation
and gives a reasonable approximation of $\chi_o(q,i\omega)$ 
in the thermodynamic limit. Our embedding procedure is based
on the observation that, in the $1\over2$-filled limit
at large $U$, $U\sim 8t$, and low temperature $T$,
the single-layer density correlation function
$\chi_o(r,i\omega)$ in $r$-space is actually of quite short range
and, already on rather small lattices, exhibits very little finite-size
dependence. Physically, this is simply a manifestation of the fact that 
the charge excitation spectrum exhibits a large Mott-Hubbard gap, 
$\Delta_{\rm MH}\sim U\gg t,T$. The Mott-Hubbard gap not only suppresses all charge
correlations; it also endows them with a very short charge correlation
length, $\xi$. In fact, from our QMC data taken on $8\times8$ lattices at $T=0.1 t$
we estimate $\xi$ to be substantially less than the in-plane lattice constant $a$.
 
Based on this observation, we approximate $\chi_o^{(L)}(r,i\omega)$ 
for a large $L\times L$ square lattice by the QMC result 
$\chi_o^{(\rm QMC)}(r,i\omega)$, obtained on a smaller lattice of size $L_o\times L_o$, 
for those $r$-vectors which fall within the (properly symmetrized)
boundaries of the smaller lattice; for r-vectors outside of 
those small-lattice boundaries, we simply set $\chi_o^{(L)}(r,i\omega)$
to zero. Specifically, for 2D lattice vectors $r\equiv(m_1 a,m_2 a)$
on the $L\times L$ ``host'' lattice,  
with integer $m_1, m_2$ and $|m_1|,|m_2|\le L/2$, we set
\begin{equation}
\chi_o^{(L)}(r,i\omega)=w(r,L_o)\chi_o^{(\rm QMC)}(r,i\omega)
\label{eq:chi_o-emb}
\end{equation}
where, for even $L_o$,
\begin{equation}
w(r,L_o)=
\left\{
 \begin{array}{ll}
             1& 
             \textrm{ if $|m_1| < L_o/2$  and $|m_2|<L_o/2$}\\
             {1\over2} &
             \textrm{ if $|m_1|=L_o/2$ and $|m_2|<L_o/2$}\\
             {1\over2} &
             \textrm{ if $|m_1|<L_o/2$ and $|m_2|=L_o/2$}\\
             {1\over4} &
             \textrm{ if $|m_1|=L_o/2$ and $|m_2|=L_o/2$}\\
             {0} &
             \textrm{ otherwise \ .}
 \end{array} 
\right.
\label{eq:w-emb}
\end{equation}
In principle, we could also set the embedding weight factors $w(r,L_o)$
to zero for those lattice vectors which fall ``on the boundary'' ({\it i.e.}
for $|m_1|=L_o/2$ or $|m_2|=L_o/2$), without noticeably changing our final
results, since even for our small QMC lattices, 
our $ \chi_o^{(\rm QMC)}(r,i\omega)$ are already
negligibly small on the boundary. 
With the above choice of $w$, Eq.~(\ref{eq:w-emb}),
we are ensuring that the large-lattice $\chi_o^{(L)}$ provides a ``natural''
interpolation of the small-lattice $\chi_o^{(\rm QMC)}$ in $q$-space,in the
sense that, after Fourier transform, identically
\begin{equation}
\chi_o^{(L)}(q,i\omega) = \chi_o^{(\rm QMC)}(q,i\omega)
\label{eq:chi-interpol}
\end{equation}
for those discrete 2D $q$-points 
$q\equiv(2\pi/a)(p_1/L_o,p_2/L_o)$,
with integer $p_1,p_2$, which
lie on the discrete $q$-grid of the smaller $L_o\times L_o$ lattice.

An additional complication in extracting $\widetilde{\epsilon}_{\rm ex}$
from the QMC data arises from the fact that the QMC simulations
are performed at finite temperature where the system exhibits
a small, but non-zero, concentration of electron
and hole charge carriers due to thermal excitation  
across the Mott-Hubbard gap. In our QMC simulations
in the grand-canonical ensemble, these thermally excited
carriers manifest themselves by the fact that 
\begin{equation}
\chi_o^{(T)}\equiv\chi_o^{(L)}(q=0,i\omega=0)>0
\label{eq:chi-T}
\end{equation}
{\it i.e.} at finite $T$ 
there is a finite thermal fluctuation in the system's
total particle number. From QMC simulations at different temperatures,
we have verified that this thermal carrier 
contribution to $\chi_o$ does indeed exhibit the expected activated
behavior, {\it i.e.}, roughly $\chi_o^{(T)}\sim \exp(-\Delta_{MH} /T)$.
If included in the calculation of $\chi(q,i\omega)$
this finite $q=0$ charge fluctuation would give rise to a 
``quasi metallic'' singular contribution
to the scalar dielectric function which would completely dominate
the $q\to0$ limit, with
\begin{equation}
\epsilon_{\rm ex}(q,i\omega=0) \sim {4\pi e^2\over {\cal V}_{\rm c} q^2}\chi_o^{(T)}\ .
\label{eq:e_ex-quasi-metal}
\end{equation}
Physically this reflects the fact that the thermally excited charge carriers are
screening out any macroscopic  Coulomb field with a finite screening
length, which is simply a manifestiation of the general principle that
no physical system can be a true insulator at finite temperature. 

To estimate the ($T=0$) insulator contribution to $\chi_o$, 
we thus subtract out the thermally activated carrier contribution
and use as our $\chi_o$-input into Eq.~(\ref{eq:p_o}) 
\begin{equation}
\chi_{\rm{o,ins}}^{(L)}(q,i\omega) = \chi_o^{(L)}(q,i\omega) 
                                  - \chi_o^{(T)}\delta_{i\omega,0} 
\label{eq:chi-ins}
\end{equation}
to calculate $P_o(q,i\omega)$. With this subtraction, the resulting
$\chi$ from Eq.~(\ref{eq:chi}) indeed, by construction, acquires 
the correct ``insulating'' long-wavelength
behavior to give a finite $\epsilon_{\rm ex}$ for $|q|\to 0$.

Taking the limit $L\to\infty$ on $\chi_{\rm{o,ins}}^{(L)}$
and then expanding the resulting $\chi_{\rm{o,ins}}^{(\infty)}$ 
around $q=0$, we can write
\begin{equation}
\chi_{\rm{o,ins}}^{(\infty)}(q,i\omega=0)=\hat{e}\cdot\widetilde{B}\cdot\hat{e}|q|^2
+{\cal O}(|q|^4)
\label{eq:chi-ins-q2}
\end{equation}
for $q\equiv|q|\hat{e}$ with unit vector $\hat{e}$ and with 
the cartesian components of the $\widetilde{B}$-tensor given by
\begin{equation}
B_{\alpha\beta} = {1\over2} \lim_{|q|\to 0}
                  {\partial^2\chi_{\rm{o,ins}}^{(\infty)}(q,i\omega=0)
                  \over
                  \partial q_\alpha \partial q_\beta}
\ \  \ \ \ \ \ \ {\rm for}\ \ \ \alpha,\beta=1,2,3\ .
\label{eq:b-tensor}
\end{equation}
Inserting $\chi_{\rm{o,ins}}^{(\infty)}(q,i\omega=0)$ from Eq.~(\ref{eq:chi-ins-q2})
into Eq.~(\ref{eq:p_o}), combining with Eqs.~(\ref{eq:chi},\ref{eq:e_ex-q}),
taking the limit  $|q|\to 0$ in  Eq.~(\ref{eq:e_ex-q}) and
comparing to Eq.~(\ref{eq:e_ex-tensor}), we get
\begin{equation}
\widetilde{\epsilon}_{\rm ex}=\epsilon_B\widetilde{1} + {4\pi e^2\over {\cal V}_{\rm c}}\widetilde{B}
\ .
\label{eq:e-b-tensor}
\end{equation}
With the cartesian coordinate axes chosen along the conventional tetragonal
symmetry directions, only the diagonal components of $\widetilde{B}$ and
$\widetilde{\epsilon}_{\rm ex}$ are non-zero with the in-plane
component given by
\begin{equation}
\epsilon_\parallel\equiv \epsilon_{{\rm ex},11}= \epsilon_{{\rm ex},22}
= \epsilon_B + {4\pi e^2\over {\cal V}_{\rm c}}B_{\parallel}\ .
\label{eq:e-para}
\end{equation}
Since, in our approximation, 
we do not include any interlayer charge correlations
in $\chi_o$, the out-of-plane (``$c$-axis'') component of $\widetilde{B}$ 
vanishes and we get
\begin{equation}
\epsilon_\perp\equiv \epsilon_{{\rm ex},33} = \epsilon_B \ ,
\label{eq:e-perp}
\end{equation}
{\it i.e.} the Hubbard electron system does not contribute to the
$c$-axis dielectric screening.

{}From our QMC results for $\chi_o^{(QMC)}$ 
on a $L_o\times L_o=8\times8$ lattice,
embedded in large $L\times L$ lattices with $L$ up to $512$,
we estimate $B_\parallel$, using Eq.~(\ref{eq:b-tensor}),
and from it the Hubbard electron system's contribution
to the in-plane dielectric constant 
\begin{equation}
\Delta\epsilon\equiv \epsilon_\parallel - \epsilon_\perp 
= {4\pi e^2\over {\cal V}_{\rm c}}B_{\parallel}\ .
\label{eq:delta-e}
\end{equation}
Using the pure on-site Hubbard potential, Eq.~(\ref{eq:v_o0}),
as our interaction potential $V_o$ in the QMC Hamiltonian,
with the standard Hubbard model parameters
stated in the previous section, we find
$\Delta\epsilon \cong 0.70$ from QMC data obtained $T=0.333t$
and $\Delta\epsilon \cong 0.93$ from QMC data obtained $T=0.1t$.
Note that this value of $\Delta\epsilon$ is independent 
of $\epsilon_B$, since $\chi_o$ and $B_\parallel$
depend only on the QMC Hamiltonian $H_o$ which 
is independent of $\epsilon_B$ for the pure Hubbard case. 

The slight ($\sim25\%$) $T$-dependence of the foregoing $\Delta\epsilon$
result suggests that the thermal fluctuations, due to
thermally excited carriers, are still noticeably 
(but not substantially) affecting the ``insulating'' charge 
correlations in the Mott-Hubbard insulator itself, at $T=0.333t$. 
This is not entirely surprising since, {\it e.g.}, 
the thermal carrier subtraction $\chi_o^{(T)}$ at that $T$
is about $10-20\%$ of typical $\chi_o$-values at 
typical non-zero $|q|\sim\pi/a$, {\it e.g.} 
$\chi_o^{(T)}/\langle \chi_o(q,i\omega=0)\rangle_{BZ}\cong 0.15 $
where $\langle ...\rangle_{BZ} $ denotes the Brillouin
zone average over $q$.
However, it is reasonable to assume that the $T=0.1t$ result 
approximates the $T=0$ limit to within a percent  or better,
since the thermal carrier subtraction 
$\chi_o^{(T)}$ at $T=0.1t$ is already 5 orders
of magnitude smaller than, say, typical $\chi_o$, {\it e.g.},
$\chi_o^{(T)}/\langle \chi_o(q,i\omega=0)\rangle_{BZ}\cong2\times 10^{-5}$.

To explore the effects of including renormalizations
due to extended Coulomb interactions in the polarization 
insertion $P_o$, we have also performed simulations for 
a $1\over2$-filled 1st neighbor extended Hubbard model,
Eq.~(\ref{eq:v_o1}), with $V_1=t=0.35{\rm eV}$ at $T=0.333t$.
The resulting change in $\Delta\epsilon$, 
$
\delta\Delta\epsilon
\equiv\Delta\epsilon|_{V_1=t}-\Delta\epsilon|_{V_1=0}
\cong-0.014\ ,
$
is only about $2\%$ of $\Delta\epsilon$.
Unfortunately, minus problems in the QMC simulations
prevent us from extending these $V_1$-dependent studies to lower
$T$ and/or more realistic larger $V_1$-values, with, say, $V_1=2-3\times t$.
However, the foregoing result clearly suggests that the extended Coulomb
effects on the polarization insertion are entirely negligible,
at least as far as the long-wavelength limit 
at $1\over2$-filling is concerned.

{}The measured values for the dielectric tensor $\widetilde{\epsilon}_{\rm ex}(\omega)$
of undoped $La_2CuO_4$ in the static limit $\omega\to0$ are
$
\epsilon_{0,\parallel}\equiv\epsilon_{\parallel}(\omega=0)
\cong 30\pm 3
$
for the in-plane component, and 
$
\epsilon_{0,\perp}\equiv\epsilon_{\perp}(\omega=0)
\cong 25\pm 3 
$ for the $c$-axis component.\cite{eps_expa}
However, these values include a large, in fact, 
dominant phonon contribution.\cite{emin}

The purely electronic contribution to the dielectric 
screening is observed at frequencies $\omega_\infty\sim 0.5-1$eV,
which are well above the phonon spectrum $\Omega_{\rm ph}\lesssim 0.1$eV, 
but at the same time still well below the electronic Mott-Hubbard charge gap
$\Delta_{MH}\sim 1.5-2{\rm eV}$. In this frequency regime, one finds
approximately frequency independent values of
$
\epsilon_{\infty,\parallel}
\equiv\epsilon_{\parallel}(\omega_\infty)\cong 4.73
$
for the in-plane component, and 
$
\epsilon_{\infty,\perp}
\equiv\epsilon_{\perp}(\omega_\infty)\cong 4.56 
$
for the $c$-axis component.\cite{eps_expb} 
The corresponding anisotropy
$
\Delta\epsilon_\infty=
\epsilon_{\infty,\parallel} - \epsilon_{\infty,\perp}\cong 0.2
$
is substantially smaller than our estimated value $\Delta\epsilon=0.9$.
However, in our model estimate, we are assuming
an isotropic background $\epsilon_B$, whereas, in the real material, 
the background itself could also be anisotropic. If we generalize our model
to allow for such a background anisotropy, we can estimate the background
in-plane and $c$-axis components, $\epsilon_{B\parallel}$ and 
$\epsilon_{B\perp}$, for our model from the experimental
$\omega_\infty$-data and our calculated $\Delta\epsilon$. Namely, 
from the generalized Eqs.~(\ref{eq:e-para},\ref{eq:e-perp},\ref{eq:delta-e}),
$
\epsilon_{B\parallel}=\epsilon_{\infty\parallel}-\Delta\epsilon
\cong 3.81
$
and
$
\epsilon_{B\perp}=\epsilon_{\perp}\cong4.57\ ,
$
corresponding to a background anisotropy 
$
\Delta\epsilon_B\equiv
\epsilon_{B\parallel}-
\epsilon_{B\perp}\cong -0.8\ .
$

One of the main conclusions from the foregoing analysis is that
the background degrees of freedom dominate the dielectric
constant at $1\over2$-filling, with the Hubbbard "conduction band"
electrons contributing only about $25\%$ to the total
electronic dielectric constant,
as observed in the $\omega_\infty$ frequency range.
The other main conclusion is that
both in the  background dielectric screening and in the
full dielectric screening (including Hubbard electrons),
the anisotropy is insignificant, compared to the actual
values of the dielectric tensor components. This conclusion
holds equally well for the purely electronic contribution to
the dielectric screening  and for the phonon contribution,
as evidenced by the nearly identical values of 
$\epsilon_{0\parallel}\sim 30$ and $\epsilon_{0\perp}\sim 25$ 
in the static limit. We will therefore in the following
continue to work with an isotropic background model, with the 
electronic ($\omega_\infty$) value
$
\epsilon_B=\sqrt{
\epsilon_{B\parallel}
\epsilon_{B\perp}
}\cong 4
$
and the analogously obtained static ($\omega\to 0$)  value 
$
\epsilon_B\cong 27
$,
providing, respectively, reasonable lower and upper limits 
for $\epsilon_B$.

{}From the estimated $\epsilon_B\!\cong\!4$ (without phonons)
or even $\epsilon_B\cong 27$ (including phonons), one obtains
substantial 1st neighbor repulsion strengths of
$V_1\equiv e^2/(\epsilon_B a)\cong0.95{\rm eV}\cong 2.7t$ 
in the former and $V_1\cong0.14{\rm eV} \cong 0.4t$
in the latter case. Thus, the extended part of the Coulomb potential, 
$V_{\rm e}(r)$, appears to be indeed strong enough,
that it could severely suppress extended (1st neighbor)
pairing potentials which are commonly invoked in both
phenomenological \cite{pines,hbs-mnn,pines-v1} 
and microscopic\cite{dagotto-rmp,bulut1,bulut2,t_j_model}
scenarios of $d$-wave pairing.\cite{esb-comnt}
It is therefore of considerable interest to find out
how this bare Coulomb interaction potential 
between the Hubbard electrons
is modified due to the metallic screening generated 
by the doped Hubbard electron system itself. 
In the next section, we will turn to this question.
%
%
\begin{figure}
\hskip5mm \epsfig{file=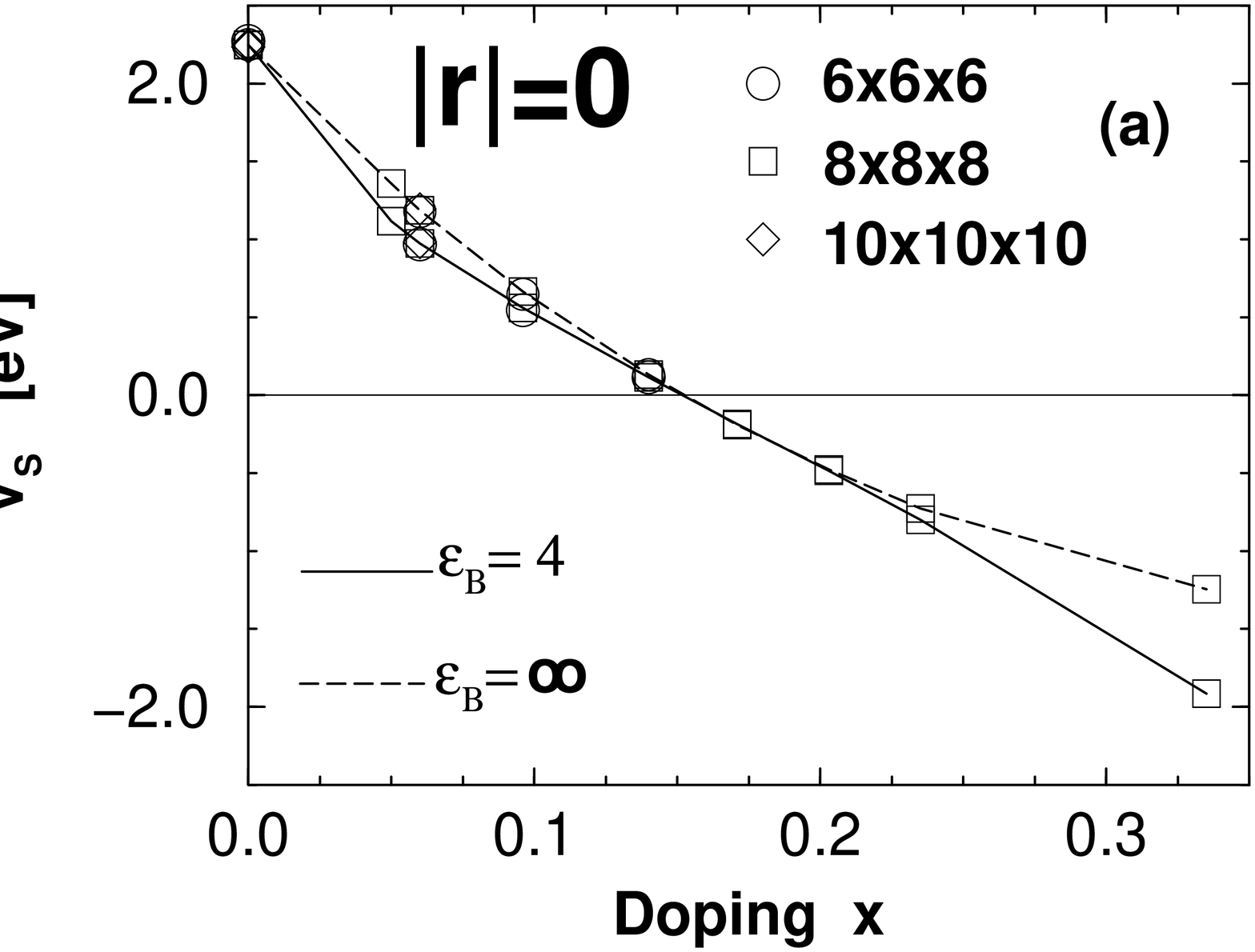,angle=-90,width=7.0cm}  \vskip0.0mm

\hskip5mm \epsfig{file=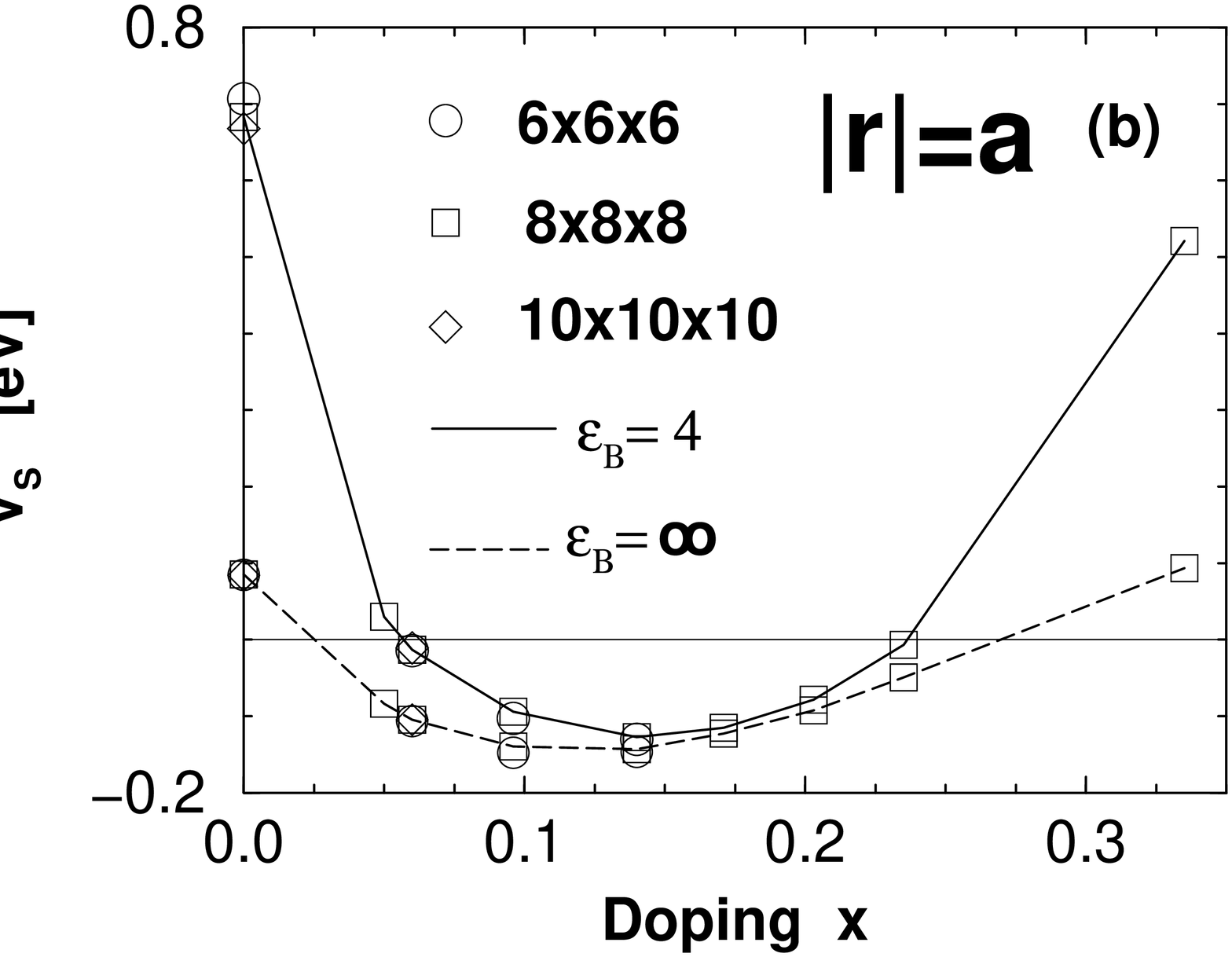,angle=-90,width=7.0cm}  \vskip0.0mm

\hskip5mm \epsfig{file=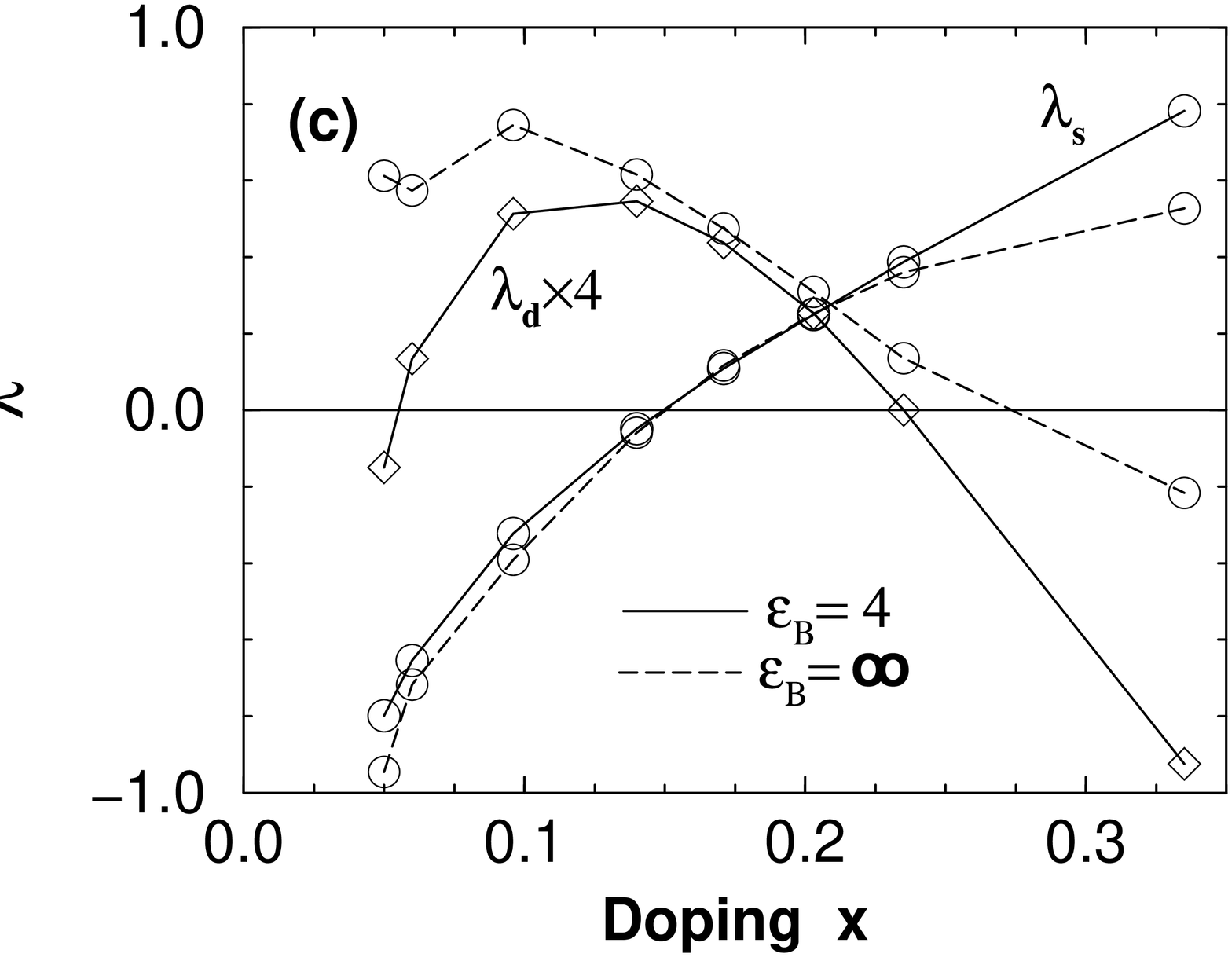,angle=-90,width=7.0cm}  \vskip0.0mm
\caption{
%
%
Screened Coulomb potential
$V_{\rm S}(r)\!\equiv\!V_{\rm S}(r,i\omega=0)$ at
(a) on site ($r=0$) and (b) at in-plane 1st neighbor
lattice vectors $r$ and (c) Eliashberg $\lambda$-parameters
for in-plane 1st neighbor $d_{x^2-y^2}$ and
on-site/1st neighbor $s$-wave pairing, all plotted
vs. hole doping concentration $x=1-\langle n_j \rangle$
at $\beta t\equiv t/T=3.0$ and $\Delta\tau t=0.0375$
for $\epsilon_B=4$ (full lines) and $\epsilon_B=\infty$ (dashed lines).
In (a) and (b), results are for $La_2CuO_4$ with $bct$ lattices 
of sizes
$6\times6\times6$, $8\times8\times8$ and $10\times10\times10$,
with estimated QMC statistical uncertainties in $V_{\rm S}$ of less than
$0.01{\rm eV}$.
In (c), results are based on $8\times8\times8$ $bct$ lattice data for $V_{\rm S}$,
extracted from $8\times8$ QMC data for $\chi_o$.
$\lambda_d$ has been multiplied $\times\!4$ for display.
%
%
}
\label{fig:v_s-x}
\end{figure}
%
%
%
\section{QMC results for $V_{\rm S}$}
\label{sec:v_screened}
In Fig.~\ref{fig:v_s-x}(a) and (b), we show results for
$V_{\rm S}(r,i\omega)$ at $i\omega\!=\!0$,
plotted {\it vs.} doping $x\equiv 1-\langle n_j\rangle$ for 
the on-site ($r\!=\!0$) and in-plane 1st neighbor $r$-vector
in the quasi-2D extended Hubbard model
on the $La_2CuO_4$ $bct$ crystal structure with $\epsilon_B=4$ and
$\epsilon_B=\infty$. Note that the latter $\epsilon_B$
represents just the pure 2D Hubbard model, with the
extended part of the Coulomb potential, $V_{\rm e}$, set to zero.
The results for $V_{\rm S}$ were extracted from QMC data for the
single-plane density correlation function $\chi_o$ of the pure
2D Hubbard model, {\it i.e.},  with $V_o$ given by Eq.~(\ref{eq:v_o0}),
as explained in Section~\ref{sec:model}, and with $V_{\rm S}(r,i\omega)$ obtained from
$V_{\rm S}(q,i\omega)$ by Fourier transform over the finite-lattice
1st Brillouin zone.
Since we are only interested in the short-distance behavior
of $V_{\rm S}$ here, we have not attempted to embed our 2D $\chi_o$-data
in larger 2D lattices, as was done in the previous Section for
the insulating dielectric function calculations. Rather, all results
shown are for $L_o\times L_o \times L_o$ 3D lattices, using QMC data for $\chi_o$
obtained on corresponding $L_o\times L_o$ 2D lattices, with $L_o=6$, $8$,
and $10$.

As shown in Fig.~\ref{fig:v_s-x}(b)
a small amount of doping
suppresses the extended $1/|r|$-repulsion for 
$r\neq0$ and causes a sign change in the 1st neighbor
and (not shown) in the 2nd and 3rd neighbor screened
potential. Thus, $V_{\rm S}(r)$ at short-range in-plane distances $r\neq0$
becomes attractive, for $x$ of order $5\%$, 
the attraction strength reaches a maximum at 
$x\!\sim\!10-14\%$, and $V_{\rm S}(r)$ turns repulsive again
at $x\!\sim\!23-28\%$.

As shown in Fig.~\ref{fig:v_s-x}(a), the on-site 
($r\!=\!0$) potential, while largely
unaffected by screening at $x=0$, 
is also rapidly suppressed with increasing $x$ 
and it also becomes attractive at larger doping,
near $x\!\cong\!15\%$. Over the doping range studied,
the screened on-site potential varies monotonically
with $x$, in contrast to the extended ($r\neq0$)
part of $V_{\rm S}$.

As a function of $\epsilon_B$, $V_{\rm S}$ varies
generally monotonically, in the directions
indicated in Figs.~\ref{fig:v_s-x}(a) and (b).
If we increase $\epsilon_B$ from $4$
to the upper estimated value of $\epsilon_B=27$,
we find that the resulting $V_{\rm S}$ will be within
$1-2\%$  of the $\epsilon_B=\infty$ (pure 2D Hubbard)
results shown in Figs.~\ref{fig:v_s-x}(a) and (b).
Overall, in comparing the quasi-2D extended Hubbard model
with $\epsilon_B=4$ to the pure 
2D Hubbard model ($\epsilon_B=\infty$),
we note that all the foregoing results are qualitatively
unaffected by the extended Coulomb repulsion. The primary
effect of the extended Coulomb terms is to move
the 1st neighbor $V_{\rm S}$ slightly in the repulsive direction, thereby
suppressing the maximal 1st neighbor attraction strength
in Fig.~\ref{fig:v_s-x}(b) by less than $10\%$.
In addition, the extended Coulomb interaction
shifts the ``optimal'' doping, where the 1st neighbor attraction maximum
occurs, from $x\cong 10\%$ in the pure 2D Hubbard model to $x\cong13\%$
in the quasi-2D extended Hubbard model with $\epsilon_B=4$.
The screened on-site potential is shifted 
slightly in the attractive direction, but, again, for doping
concentrations up to $20\%$, the effect is typically smaller than $10\%$.

These results suggest that, despite the substantial strength of the
bare extended Coulomb potential $V_{\rm e}$ at near-neighbor distances, 
the local screened potential in the Hubbard electron system at finite
doping is largely unaffected by extended Coulomb interactions. 
Recall from the previous section that the 
Hubbard electrons are largely ineffective in
providing dielectric screening in the $1\over2$-filled insulator.
By contrast, the results in Fig.~\ref{fig:v_s-x} suggest that the Hubbard
electrons' metallic screening is very strong, in fact,
strong enough to completely overwhelm the extended 
Coulomb repulsion, once a sufficient amount
of doping and sufficient electronic background screening, 
with $x>5\%$ and $\epsilon_B\gtrsim4 $, say, is present. 
Additional background screening due to phonons will
not have any significant effect on $V_{\rm S}$ under these
conditions.

>>From the foregoing discussion, it also becomes clear that the
presence and frequency dependence of the phonon contribution
to $\epsilon_B$ will not substantially affect the
Hubbard electrons' metallic screening at finite doping.
If one were to include a phononic frequency dependence in
$\epsilon_B$, with a typical phonon frequency scale
$\Omega_{\rm ph}\ll 8t$, then as a function of $i\omega$,
one would find that $V_{\rm S}(r,i\omega)$ is given roughly
by the pure Hubbard result 
({\it i.e.} the dashed lines in Fig.~\ref{fig:v_s-x}) 
at frequencies $|i\omega|\ll \Omega_{\rm ph}$ 
where the phonons contribute to the background screening;
and $V_{\rm S}(r,i\omega)$ is given roughly 
by the extended Hubbard result with $\epsilon_B\cong 4$
({\it i.e.} by the full lines in Fig.~\ref{fig:v_s-x}) 
at frequencies $|i\omega|\gtrsim \Omega_{\rm ph}$ 
where the phonons do not contribute to the background screening.
Over most of the doping range of interest, {\it e.g.},
between $x\sim10\%$ and $x\sim20\%$, this additional
``phononic'' frequency dependence of $V_{\rm S}$, for $|i\omega|$ on the 
scale of $\Omega_{\rm ph}$, is thus quite small, since the 
$\epsilon_B=4.0$ extended Hubbard results are not very different
the $\epsilon_B=\infty$ pure Hubbard results for $V_{\rm S}$.
More importantly, the qualitative features of $V_{\rm S}$,
such as its doping dependence and the overscreening effects
at finite doping will not be changed by the phononic effects.
The foregoing argument implicitly assumes
that the characteristic frequency scale of the
electronic density correlations, {\it i.e.}, of $\chi_o(q,i\omega)$,
is much higher than $\Omega_{\rm ph}$. From earlier QMC studies, it appears that
this condition is indeed satisfied, since the
spectral weight for density fluctuations in the Hubbard
model extends over a frequency range comparable to the electronic
bandwidth.\cite{preuss} 

By the same arguments, the phononic frequency dependence 
of $V_{\rm S}$ will become increasingly important at low doping 
concentrations where the metallic screening of the
Hubbard electrons is too weak to suppress the
extended Coulomb effects. This observation may offer a
possible explanation for the doping dependence of the
isotope effect in the superconducting transition
temperature of the cuprates. The observed isotope
exponent\cite{franck} $\alpha$ is generally a decreasing 
function of doping, in the underdoped regime,
and becomes negligibly small (compared to the classical
BCS value of $\alpha={1\over2}$) in the ``optimal'' doping 
range near $x\sim15\%$. This behavior is qualitatively
consistent with the notion that, with increasing
doping, the phonon effect becomes less
and less important in the screened potential.
Along with its doping dependence,
also the overall magnitude of the isotope
exponent in underdoped cuprates (which can exceed
the classical BCS value\cite{franck}) has been a long-standing
theoretical puzzle,\cite{hbs-chp} given the assumption of an electronic
pairing mechanism. It remains to be explored within specific
pairing models whether
the phononic frequency dependence of the screened Coulomb
potential is large enough to also explain the
observed magnitudes of $\alpha$
in the underdoped regime. 

Minus sign problems at finite doping 
unfortunately limit our simulations to $T\!\geq\!0.33t$.
However, at least in that temperature regime, we find $|V_{\rm S}|$ 
to be increasing with decreasing $T$. This suggests that the
overscreening effects ({\it i.e.} the attraction in $V_{\rm S}<0$) 
becomes stronger than shown in Fig.~\ref{fig:v_s-x}
at lower $T$.

The presence of {\it both} a strong Hubbard-$U$ 
{\it and} finite doping density $x\!>\!0$
are crucial for these overscreening effects
to arise. If we replace the fully $V_o$-renormalized 
$P_o$ in (\ref{eq:v_s}) by, say, the non-interacting 
("RPA") polarization bubble $P_{RPA}$, we also
obtain a suppression of $V_{\rm S}(r)$. However, both the
on-site ($r\!=\!0$) and the short-range
extended part ($r\neq 0$) of $V_{\rm S}$ 
remain repulsive in RPA over the whole doping range
studied here.\cite{liu_levin}

In the undoped large-$U$ system, $V_{\rm S}(r)$
is reduced relative to $V(r)$, by a roughly 
$r$-independent factor comparable to the ratio
$\epsilon_{\rm ex}/\epsilon_B$,
for $r\neq0$, {\it  i. e.} $V_{\rm S}$ is reduced relative to $V$ 
but retains a repulsive $1/|r|$-dependence. 
This is expected for the screening of a $1/|r|$ potential
in an insulator and confirms the insulating character
of the ${1\over2}$-filled Hubbard system.
The $1/r$ long-distance behavior of $V_{\rm S}(r)$ can of course again
be traced back to the fact that long-wavelength charge
fluctuations are suppressed in the $1\over2$-filled
insulating state. For $x=0$, $\chi(q,i\omega)\sim|q|^2\to 0$, 
and hence also $P(q,i\omega)\sim|q|^2\to 0$ for $|q|\to 0$.
As a consequence, the screening denominator $1-V(q)P(q,i\omega)$
in Eq.~(\ref{eq:v_s}) remains finite for $q\to 0$ and $V_{\rm S}(q,i\omega)$
inherits the $1/|q|^2$ singularity of $V(q)$, thus 
its $1/|r|$-dependence upon Fourier transforming back to $r$-space.

By contrast, in the doped system,
the screening takes on a noticeably metallic character and 
the $r$-dependence of $V_{\rm S}$ changes dramatically, even
on the finite lattice sizes and high temperatures 
accessible with our QMC approach.
Already at small doping concentrations, $x\gtrsim5\%$,
$V_{\rm S}(r,i\omega)$ dies out much faster with $|r|$ than in the 
undoped case. For example, at ``optimal doping'' where
the 1st neighbor $V_{\rm S}$ is maximally attractive,
the in-plane 2nd neighbor $V_{\rm S}$ [at $r=(a,a,0)$]
is about 4 times smaller in magnitude than the 
in-plane 1st neighbor $V_{\rm S}$ [at $r=(a,0,0)$]. 

The metallic screening also strongly suppresses
the inter-layer Coulomb repulsion, with $V_{\rm S}$
at the nearest inter-layer distance $r=(a/2,a/2,d)$
being reduced by a factor of probably more than 
500 relative to the bare $V$ at that distance 
and by a factor of more than 100 relative to the 
in-plane 1st neighbor $V_{\rm S}$.
Notice here that the quasi-2D Hubbard electron
system {\it can} produce inter-layer screening
even though our model does {\it not} include
any matrix elements for inter-layer electron transfer. 
Unlike the in-plane components of $V_{\rm S}$, 
its inter-layer components remain repulsive
at finite dopings, over the doping range
explored in Fig.~\ref{fig:v_s-x}, {\it i.e.},
we do not find any evidence for inter-layer
overscreening.

The essential features of $V_{\rm S}$ are robust against
substantial modifications of the extended Coulomb terms.
Note here that, in real materials, the simple $1/|r|$-dependence of
$V_{\rm e}(r)$ will of course be modified locally
by local field effects.\cite{eps_th,liu_levin}
However, our results do not change by 
more than $20-30\%$ if we in- or decrease
$V_{\rm e}(r)$ locally, at 1st, 2nd, and/or 3rd
neighbor distances, by up to $30\%$,
relative to Eq.~(\ref{eq:coulomb}),
The latter is a conservative upper limit for such
local field effects, based on the cuprates' Wannier orbital 
and crystal structure.\cite{sf,eps_th,liu_levin}
Using different layered 3D geometries, such as a
$YBa_2Cu_3O_7$ bi-layer structure, also
does not change the results in 
Fig.~\ref{fig:v_s-x} by more than a few percent.

We have also carried out QMC simulations with
the 1st neighbor extended Hubbard model, Eq.~(\ref{eq:v_o1}),
at finite doping, thereby including extended Coulomb effects in our 
in our polarization insertion $P$. Surprisingly, we find that this 
actually increases the 1st neighbor attraction of $V_{\rm S}$ and, to
a lesser extent, also the on-site attraction. For example, if we simulate
and analyze according to Eqs.~(\ref{eq:v_s},\ref{eq:p_o})
an in-plane 1st neighbor Hubbard model, with $V(r)=V_o(r)$ given by
Eq.~(\ref{eq:v_o1}) and a 1st neighbor repulsion strength $V_1=0.5t=0.175{\rm eV}$, 
then the magnitude of the 1st neighbor attraction in $V_S(r,i\omega=0)$
at a near-optimal doping of $x=14\%$ increases by about $16\%$ relative 
to the pure on-site Hubbard model results shown in Fig.~\ref{fig:v_s-x}(c).
Notice that this is a comparison of two exact results, since
we have used the same potential $V(r)$ in Eq.~(\ref{eq:v_s})
as was used as our QMC potential $V_o(r)$, 
both in the pure Hubbard and in the 1st neighbor extended Hubbard
calulation, {\it i.e.} $\chi_o=\chi$ and $P_o=P$ exactly.
The foregoing result implies that the 1st neighbor overscreening
(in the $5-25\%$ doping range) is not only robust against
extended Coulomb interactions, but, in fact, could be enhanced
if extended Coulomb interactions are included exactly in the
polarization insertion. This result also suggests the interesting
possibility that extended Coulomb interactions may actually contribute
constructively to the pairing attraction for extended
superconducting pair wavefunctions,\cite{leggett}
and specifically to $d_{x^2-y^2}$-pairing,
as discussed in Section~\ref{sec:pairing}.
The effects of including extended Coulomb interactions 
in the polarization insertion need to be investigated further.
Specifically, the effects of longer-range Coulomb terms
(2nd, 3rd, ...) neighbor need to be explored.

Only at unphysically large $V_1$, 
{\it i.e.} unphysically small $\epsilon_B$,
does our approach break down, due to charge density wave instabilities.
These instabilities are signaled 
by $1/\chi(q,i\omega\!=\!0)\!\to\!0$ at some point
in $q$-space. For the parameter range explored in the
present paper, this occurs only for $\epsilon_B\lesssim 2.0$.

\section{Exact proof of on-site overscreening 
and its implications for the polarization insertion}
\label{sec:ex_proof}
In order to see why a sufficiently large $U$ at finite 
doping $x\equiv 1-\langle n_j\rangle$ 
{\it must} cause an on-site overscreening effect, 
we consider first the pure Hubbard model, 
$V(r)\equiv U\delta_{r,0}$,
where, from the {\it exact} Eq.~(\ref{eq:v_s}),
\begin{equation}
V_{\rm S}(r,i\omega)=U\delta_{r,0}-U^2\chi(r,i\omega) \ ,
\label{eq:v_s-hub}
\end{equation}
with $V_{\rm S}(r,i\omega)$ and $\chi(r,i\omega)$ 
denoting the respective Fourier transforms
of $V_{\rm S}(q,i\omega)$ and $\chi(q,i\omega)$
back into $r$-space. Clearly, the on-site $-U^2\chi$-term
in Eq.~(\ref{eq:v_s-hub}) is attractive for all $i\omega$, since
$\chi(r\!=\!0,i\omega)$, the autocorrelation
function of $\Delta n_j$, is always positive.

At $x\!=\!0$, {\it i.e.}
in the $1\over2$-filled insulator,
charge fluctuations are suppressed
by the Mott-Hubbard gap.
For $U\to\infty$, one finds from a large-$U$ expansion
that $\chi(r,i\omega=0)\!\sim\!{\cal O}(t^2/U^3)$,
and, in Eq.~(\ref{eq:v_s-hub}),
$V_{\rm S}(r\!=\!0,i\omega\!=\!0)\!\cong\!U-{\cal O}(t^2/U)>0$
which remains repulsive for large $U\gg t$. 
The crucial point here is that, 
at finite $|x|\!>\!0$, even an infinitely
large $U$ can {\it not} completely suppress 
the charge fluctuations, since 
$n_j$ is not conserved at finite $|x|$,
regardless of $U$. Specifically, the on-site
component $\chi(r\!=\!0,i\omega)$ approaches a positive,
{\it non-zero} limit [of ${\cal O}(x/t)$, 
by a simple $U\!=\!\infty$ scaling argument] 
for $U\to\infty$. 
Hence, the $-U^2\chi$-term will 
overcome the bare $U$-term in 
(\ref{eq:v_s-hub}) and $V_{\rm S}(r\!=\!0,i\omega)$ 
must become attractive, {\it i.e.} on-site
overscreening must occur, for sufficiently large $U$.
Also, as a consequence, for sufficiently large $U$, 
$V_{\rm S}(r\!=\!0,i\omega\!=\!0)$
{\it must} change sign as a function of $x$.

Formally, the suppression of $\chi$ at 
$1\over2$-filling arises in the strong-coupling expansion
because the large repulsive $U$ effectively projects out the
low-energy Hilberspace sector containing only states without doubly
occupied sites. Upon projection onto the low-energy sector, 
the on-site charge operators $n_j$ become exactly conserved 
with $n_j=1$ and $\Delta n_j=0$ 
at all sites for $U\to\infty$, giving $\chi\equiv 0$ from Eq.~(\ref{eq:chi_o}).
At large, but finite $U$, intersite charge transfer processes
via the hybridization term $H_t$, must necessarily go through 
``virtual'' intermediate states containing at least one 
doubly occupied site. To 2nd order in the hybridization $H_t$,
each of the two matrix elements exciting from the low-energy sector
into this high-energy sector and back is of order $t/U$ while the
inverse energy denominator associated with the virtual excitation is
of order $1/U$, resulting in $\chi\sim t^2/U^3$.
At finite doping, on the other hand, the projected $n_j$ are not conserved.
Even for $U=\infty$, the doping induced holes can still 
move through the lattice, via the $H_t$-term, since the
intersite charge transfer can proceed without having to go through
high-energy intermediate states involving double occupancy. Hence,
$\chi$ approaches a non-zero limit, $\chi_\infty\neq0$, at finite doping concentration for
$U\to\infty$.

Note that the foregoing large-$U$ argument constitutes an {\it exact} analytical 
proof of on-site overscreening in the asymptotic limit $U\to\+\infty$. 
The proof holds both on finite and infinite lattices in any spatial 
dimension and it immediately generalizes to the full extended Hubbard model, 
Eqs.~(\ref{eq:model},\ref{eq:coulomb}) since the $-U^2\chi$-term
remains the dominant screening contribution for $U/t\to\infty$ at finite $x>0$,
even in the presence of a finite extended interaction term
$V_{\rm e}\!\ne\!0$. Our QMC results in Fig.~\ref{fig:v_s-x}a 
not only confirm these exact large-$U$ results; 
they also show that the large-$U$ scenario is realized 
in the physically relevant parameter regime\cite{sf} $U\!\sim\!8-12 t$.

The occurence of on-site overscreening, or close proximity
to it, is accompanied by profound effects in the polarization insertion
$P(q,i\omega)$, namely, by $q$-dependent singularities of $P$ 
on the (analytically continued) $i\omega$-axis.
To see this, recall that $V_{\rm S}(r=0,i\omega)$ is just the Brillouin 
zone average of $V_{\rm S}(q,i\omega)$. Hence, in order to get 
$V_{\rm S}(r=0,i\omega)<0$, there must be some region in $q$-space, 
for which $V_{\rm S}(q,i\omega)<0$. On the other hand, since 
$\chi(q,i\omega)\sim{\cal O}(1/\omega^2)$ for $|i\omega|\to\infty$,
it follows from Eq.~(\ref{eq:v_s}) that $V_{\rm S}(q,i\omega)\to V(q)>0$
for $|i\omega|\to\infty$, {\it i.e.}, in the high frequency limit
the screened potential approaches the repulsive bare potential. Hence,
for those $q$ for which $V_{\rm S}(q,i\omega)<0$ at some (low) frequency
$i\omega$, there must exist at least one $i\omega^{(s)}(q)$, on 
the imaginary frequency axis, such that $V_{\rm S}(q,i\omega)$, 
analytically continued onto the continuous $i\omega$-axis, 
goes through zero, 
\begin{equation}
\lim_{i\omega\to i\omega^{(s)}(q)} V_{\rm S}(q,i\omega)=0 \ .
\label{eq:v_s-to-0}
\end{equation}
By Eq.~(\ref{eq:v_s}), this implies 
\begin{equation}
\lim_{i\omega\to i\omega^{(s)}(q)} 1/\chi(q,i\omega)= V(q)
\label{eq:inv-chi-to-v}
\end{equation}
and, by Eq.~(\ref{eq:chi}) which is equivalent to 
$1/P(q,i\omega)=V(q)-1/\chi(q,i\omega)$, this implies 
\begin{equation}
\lim_{i\omega\to i\omega^{(s)}(q)} 1/P(q,i\omega)= 0\ .
\label{eq:inv-p-to-0}
\end{equation}
In other words, $1/P(q,i\omega)$ changes sign 
and $P(q,i\omega)$ must be singular at $i\omega^{(s)}(q)$.

Note that this is clearly a strong correlation effect.
Weak coupling approximations to $P(q,i\omega)$, notably RPA, do not
give such a singularity in $P(q,i\omega)$. In RPA, 
$1/P(q,i\omega)<0$ at all $q$ and $i\omega$. It is therefore
not surprising that RPA cannot reproduce the on-site overscreening
effect. Note also that this
singularity in $P$ does not imply any singularities ({\it i.e.}
instabilities) in physically observable quantities, such as $\chi(q,i\omega)$.
In fact, $\chi(q,i\omega)$ is perfectly regular at $i\omega^{(s)}(q)$,
as implied by Eq.~(\ref{eq:inv-chi-to-v}) and $V(q)>0$.

However, the singularity of $P(q,i\omega)$ does imply singularities
of certain vertex functions to which both $P$ and the single-particle
self-energy $\Sigma$ are diagrammatically related. For example,
both $P$ and $\Sigma$ can be expressed in terms of an appropriately
defined 3-point vertex function $\Lambda(k,i\nu;q,i\omega)$,
with entering boson momentum-energy $(q,i\omega)$, 
entering fermion momentum-energy $(k,i\nu)$ (where
$i\nu$ is an odd Matsubara frequency) and exiting fermion 
momentum-energy $(k+q,i\nu+i\omega)$, such that\cite{mahan}
\begin{eqnarray}
&&P(q,i\omega)
\nonumber\\
&&=\Big({T\over N}\Big) \sum_{k,i\nu}
G(k,i\nu) G((k+q,i\nu+i\omega)
\Lambda(k,i\nu;q,i\omega)
\label{eq:p-vert}
\end{eqnarray}
and
\begin{eqnarray}
&&\Sigma(k,i\nu)
\nonumber\\
&& = \Sigma_{\rm H}(k)  
\nonumber\\
&&-\Big({T\over N}\Big)\sum_{q,i\omega} 
V_{\rm S}(q,i\omega)G(k+q,i\nu+i\omega)\Lambda(k,i\nu;q,i\omega)
\label{eq:s-vert}
\end{eqnarray}
where $G(k,i\nu)$ is the single-particle Green's function
and $\Sigma_{\rm H}$ denotes the self-energy contribution from the
Hartee diagram. Note that the singularity of $P$
implies an analogous singularity in the $i\omega$-dependence
of $\Lambda$ for $i\omega\to i\omega^{(s)}(q)$. However, in 
the foregoing expression for $\Sigma$, which is, in principle, a physically 
observable quantity, the singularity of $\Lambda$ is cancelled 
by the vanishing of the other factor in the summand,
$V_{\rm S}(q,i\omega)\to 0$ for $i\omega\to i\omega^{(s)}(q)$.
The formal implications of this singularity in $P$ and $\Lambda$
need to be further investigated.

As discussed in Section~\ref{sec:v_screened}, RPA results 
show that, for near (1st, 2nd, ...) neighbor $r$'s, 
$\chi(r,i\omega\!=\!0)$ is negative at small $U$,
whereas QMC results suggest that the near-neighbor
$\chi(r,i\omega\!=\!0)$ 
becomes positive at finite doping when
$U$ exceeds a doping dependent threshold of 
order several $t$. 
>>From Eq.~(\ref{eq:v_s-hub}) one sees
that this positive near neighbor
$\chi(r,i\omega\!=\!0)$ gives rise
to the near neighbor attraction in $V_{\rm S}$
at large $U$ and finite $x$, as
displayed in Fig.~\ref{fig:v_s-x}b. 
Hence, the QMC and RPA results taken together suggest that
near-neighbor overscreening is fundamentally 
also a large-$U$ effect, just like on-site overscreening.
However, note here that $\chi(r,i\omega)$
for $r\neq0$ is {\it not} an autocorrelation function
and its sign is allowed to be either positive or negative, depending
on the model parameters. Because of this, it does not
seem to be possible to generalize the above large-$U$ overscreening proof 
to show the existence of near- (1st 2nd, ...) neighbor overscreening 
in $V_{\rm S}(r,i\omega)$ by analytical means. Also, unlike
the on-site overscreening case, the existence of
near-neighbor overscreening does not necessarily imply
the existence of singularities in $P(q,i\omega)$.
\section{$d$- and $s$-wave pairing strengths}
\label{sec:pairing}
Given the attractive nature of the screened
potential at finite doping, it is tempting to
ask whether this attraction could give rise to superconductivity
and, if so, of what pairing symmetry.
A potential advantage of our diagrammatic expansion
in the charge representation is the large
reduction of the overall strength 
of $V_{\rm S}$ in the $10-20\%$ doping range,
compared to the bare Hubbard-$U$.
This suggests the possibility of 
carrying out controlled, self-consistent weak-coupling
expansions in which the fully screened $V_{\rm S}$, 
rather than the bare $V$ or $U$, serves as the small parameter. 
Such an expansion can be formulated diagrammatically\cite{mahan}
by retaining only skeleton diagrams in which none of the
interaction lines contain any polarization
insertions and each interaction line represents a $V_{\rm S}(q,i\omega)$. 
Superconducting instabilities can then be 
studied in terms of such a perturbative
approximation to the irreducible particle-particle
vertex, expanded to 1st order in $V_{\rm S}$.

As a first step in that direction,
we have explored possible $V_{\rm S}$-induced or -enhanced
superconducting pairing instabilities, using
the standard Eliashberg-McMillan (EM) approach.\cite{mcmillan}
A convenient measure of the pairing strength of $V_{\rm S}$
are the dimensionless EM $\lambda$-parameters,
defined in terms of the Fermi surface "expectation values" 
of $V_{\rm S}(k-k',i\omega\!=\!0)$ 
for relevant Cooper pair trial wavefunctions $\eta(k)$
in electron momentum ($k$-) space, as described, {\it e.g.}, in
Refs.~\onlinecite{hbs-mnn} and ~\onlinecite{mcmillan}. 

In Fig. 1(c), we show the EM parameters 
$\lambda_s$, for on-site $s$-wave 
(and, identically, for in-plane 
1st neighbor $s$-wave),\cite{ext-s-wave}
and $\lambda_d$, for in-plane 1st neighbor $d_{x^2-y^2}$ pairing, 
with respective pair wavefunctions $\eta_s(k)\equiv 1$ 
and $\eta_d(k)=\cos(ak_x)-\cos(ak_y)$.
To carry out the required Fermi surface integrals, our 3D
$V_{\rm S}(q,i\omega)$ was interpolated from 
the finite $8\!\times\!8\!\times\!8$ lattice $q$-grid 
onto a $200\!\times\!200\!\times\!200$ $q$-grid,
using the 3D version of the $q$-interpolation scheme
described in Section~\ref{sec:dielectric}
for the 2D $q$-interpolation of $\chi_o$.
Applying this interpolation scheme to $V_{\rm S}(q,i\omega)$
is justified here, analogous to the above $\chi_o$-interpolation,
by the fact that, at finite doping, $V_{\rm S}(r,i\omega)$ 
is of very short-range in $r$-space, 
due to the ``metallic'' character of the screening.

At low doping, the dominant 
attractive ($\lambda\!>\!0$) channel 
is $d_{x^2-y^2}$ with $\lambda_d$ reaching a maximum of
$\sim\!0.15-0.17$ near $x\!\sim 10-14\%$. 
$\lambda_s$ is repulsive at low doping, 
but becomes strongly attractive
at larger doping $x\gtrsim15\%$. 
Thus, as expected on symmetry grounds,
$\lambda_s$ and $\lambda_d$
reflect the doping dependence 
of the on-site and 1st neighbor 
attraction $V_{\rm S}$ shown in 
Fig.~\ref{fig:v_s-x}(a) and (b), respectively.
We note in passing that the doping dependence
of $\lambda_d$ is reminiscent of the observed doping dependence of 
the superconducting $T_{\rm c}$ in the cuprates.\cite{scalapino}
The $\lambda$-values for near-neighbor pair wavefunctions
of other symmetries ($p$, $d_{xy}$, $g$) are
small compared to $\lambda_s$ and $\lambda_d$
and for that reason not further discussed here. 

The spectral weight of $\chi(q,i\omega)$ extends
up to values $\Omega_\chi\!\sim\!8-10t$.\cite{preuss}
In the EM analysis,\cite{mcmillan} this "boson" energy scale,
together with $\lambda$, determines the superconducting $T_{\rm c}$, roughly as 
$T_{\rm c}\sim\Omega_\chi\exp(-1/\lambda)$.
Because of the large $\Omega_\chi$-scale, it may be
possible, at least within the EM approximation, to achieve high $T_{\rm c}$'s
even for moderate coupling values $\lambda\!<\!1$.

We should caution here that the foregoing QMC results can 
of course only suggest the general, qualitative 
trends in $V_{\rm S}$ and $\lambda$ parameters,
due to the high temperatures and small lattice 
size limitations inherent in the QMC approach. 
Also, our EM approach is based formally 
on a perturbative expansion of the
exact irreducible particle-particle vertex to 1st order in $V_{\rm S}$.  
This approximation should be relied upon,
if at all, only close to the "cross-over"
$x_C\cong 15\%$ where $V_{\rm S}(r\!=\!0,i\omega\!=\!0) = 0$,
so that $|V_{\rm S}(q,i\omega)|\ll 8t$, at least for low
frequencies $|i\omega|\ll 8t$.
i.e. roughly in the $10-20\%$ doping range.
At large over- or underdoping 
($x\gtrsim 20\%$ or $x\lesssim 10\%$),
where $|\lambda_s|\sim{\cal O}(1)$,
corrections of higher order in $V_{\rm S}$ 
are likely to contribute strongly 
to the self-energy and to the irreducible 
particle-particle vertex, thereby causing the
EM approximation, with $V_{\rm S}$ as the effective pairing potential,
to break down. The increasingly attractive $\lambda_s$-values
at $x\gtrsim 25\%$ doping, in Fig. 1c, do 
therefore not necessarily imply 
increasingly strong $s$-wave pairing tendencies.
The effects of vertex corrections, of higher order in $V_{\rm S}$,
need to be further explored, both in the over- and underdoped regimes.

It is also important to realize that the strong 
reduction in overall screened potential strength
$|V_{\rm S}|$ is only a necessary, not a sufficicient condition
for the applicability of the EM approach. Note in
particular, that $V_{\rm S}$ is ``weak'' (and, where applicable, attractive)
only at low frequencies. At large frequencies, with $|i\omega|$
of the order of the bandwidth $8t$, the charge correlation
function begins to die out, with $\chi(q,i\omega)\sim 1/|i\omega|^2$,
and, by Eq.~(\ref{eq:v_s}), $V_{\rm S}(q,i\omega)\to V(q)$ for $|i\omega|\to\infty$.
In other words, at high frequencies, the screened potential
recovers the full strength of the bare potential.
Physically, this simply reflects the fact that at high frequencies,
$|i\omega|\gg 8t$, the conduction electron system is ``too slow''
to provide a screening response to very rapidly varying
fields. Since the characteristic frequency scale of these 
charge fluctuations is quite high, of the order of 
the bandwidth,\cite{preuss} the conventional underpinnings of 
the EM approximation, such as the Migdal theorem, are 
not necessarily satisfied. 
Thus, even in the ``near-optimal'' doping regime 
where the low-frequency $V_{\rm S}$ is weak,
it needs to be re-examined whether corrections 
to the irreducible particle-particle vertex
of higher order in $V_{\rm S}$ are indeed small.

With these caveats in mind, we should compare 
our results for the pairing strengths in the Hubbard model
to earlier studies of the superconducting 
pairing correlations \cite{white,imada,moreo,hubb-cpmc}
and of the effective pairing potential \cite{bulut1,bulut2,scalapino-tjp,bulut0}
in the Hubbard model. Early studies of pair correlation functions\cite{white,imada,moreo}
and pair bound state symmetries on small lattices\cite{dagotto,parola}
using formally exact QMC simulation methods\cite{white,imada,moreo,dagotto}
and exact diagonalization,\cite{parola}
indicated a tendency towards $d_{x^2-y^2}$-pairing.
In finite-$T$ simulations,\cite{white,imada,moreo}
this was suggested by an increase in  
$d_{x^2-y^2}$ pairing correlations with decreasing temperature.
However, due to the QMC minus-sign problem, these
exact QMC studies were limited to small lattice sizes
and rather high temperatures (in physical units about 
an order of magnitude higher than the observed $T_{\rm c}$ 
scale in the cuprates). It may therefore be difficult to
draw meaningful conclusions about the low-temperature
long-range pairing correlations of the model from
the finite-system and/or finite-$T$ exact QMC data
for pair correlation functions. 

Also, by contrast, more recent studies of pairing
correlations in the groundstate of the Hubbard model,
based on the ``constrained path Monte Carlo'' (CPMC)
approach, did not provide any evidence for long-range
pairing correlations of either $d_{x^2-y^2}$- or $s$-wave
symmetry.\cite{hubb-cpmc} However, the CPMC approach is based on a
(in principle uncontrolled) variational
approximation which, albeit remarkably successful
in reproducing exactly known groundstate energies,
may not necessarily reproduce the exact pairing
correlations.
Thus, existing QMC studies of  pair correlation
functions in the Hubbard model are, at best, inconclusive
as far as $d_{x^2-y^2}$-pairing is concerned and, with all
the above-cited limitations, they do not provide
evidence for $s$-wave pairing in the near-$1\over2$-filled
Hubbard model.

In addition to the above-cited ``technical'' limitations, 
there is a further, physical reason why especially 
long-range $s$-wave pairing correlations, 
if existent in the Hubbard model, 
may escape detection in QMC simulations of
pair correlation functions.
To understand this, note that retardation 
plays a central role in the physical 
origin of the overscreening attraction in our $V_{\rm S}$.
For example, for the pure Hubbard model, Eq.~(\ref{eq:v_s-hub}),
Fourier transformed to the (imaginary) time ($\tau$) domain, becomes 
\begin{equation}
V_{\rm S}(r\!=\!0,\tau)= U\delta(\tau) - U^2\chi(r\!=\!0,\tau)\ .
\label{eq:v_s-tau}
\end{equation}
A pair of electrons, $e$ and $\bar{e}$, say, 
is subject to the instantaneous, bare repulsive 
$U$ interaction term only if they both occupy the
same site $j$ at the same time. By contrast, the attractive
screening term, $-U^2\chi$, is retarded, {\it i.e.}, physically
speaking, the charge polarization caused by $\bar{e}$ 
at site $j$ can still be felt by $e$ after a time lag
$\theta>0$, when $\bar{e}$ has already left $j$.
Thus, the on-site overscreening arises
from processes wherein the two
electrons interact on the same site, but at different times,
via their local charge polarizations, thereby evading their
bare, instantaneous repulsion.
The time scale for this retarded interaction is
governed by the frequency spectrum of the dynamical density 
correlation $\chi$, {\it i. e.}, for the relevant parameter range
it is of order of the inverse bandwidth $(8t)^{-1}$.\cite{preuss}
Pair correlations induced by such a potential, with
retarded attractive and instantaneous repulsive
contributions, should be most easily detected 
via {\it time-delayed} pair creation order parameters
of the general form
\begin{equation}
\Delta_\theta=
c^\dagger_{j,\sigma}(\theta) 
c^\dagger_{\bar{j},\bar{\sigma}}
\label{eq:delta_theta}
\end{equation}
where
$
c^\dagger_{j,\sigma}(\theta)\equiv
e^{\theta H} c^\dagger_{j,\sigma} e^{-\theta H}\ .
$
In other words, $\Delta_\theta$ creates 
the second electron $e$ with time lag $\theta$ 
relative to the first electron $\bar{e}$. 
For the on-site $s$-wave ($j=\bar{j}$) case,
one should thus consider time-delayed pair correlation 
functions of the general form
\begin{equation}
C(r_j-r_{j^{\bf\prime}},\tau,\theta,\theta^{\bf\prime})
=
\langle 
T\big[\ 
 c_{j\downarrow}(\tau+\theta) 
 c_{j\uparrow}(\tau)\ 
 c_{j^{\bf\prime}\uparrow}^\dagger(\theta^{\bf\prime}) 
 c_{j^{\bf\prime}\downarrow}^\dagger(0)\ 
\big]
\rangle
\label{eq:pair_corr}
\end{equation}
where $T[...]$ implies fermion time ordering.
The strongest signal for pairing correlations
should then be detected for typical values of 
$\theta$ and $\theta^{\bf\prime}$ which are
comparable to the retardation time scale of the
attractive part of the potential, {\it i.e.},
in the case of our $V_{\rm S}(r,\tau)$, for
$\theta,\theta^{\bf\prime}\sim(8t)^{-1}$.
By contrast, the existing QMC studies of pairing correlation functions
in the Hubbard model have so far considered only {\it simultaneous} 
pair correlations
$C(r_j-r_{j^{\bf\prime} },\tau,\theta,\theta^{\bf\prime})$
with $\theta\!=\!\theta^{\bf\prime}\!=\!0$. 
In such a simutaneous pair correlation function, the detectable signal 
for $s$-wave pairing is expected to be very weak
(and, in the case of pairing with odd frequency parity, identically zero)
if the underlying pairing potential happens to have 
a time-dependent structure of the sort described above for
$V_{\rm S}(r,\tau)$, {\it i.e.} a strong instantaneous repulsion superimposed 
on a retarded attraction. 
We note that for such a temporal structure of the
pairing potential the possibility of a triplet $s$-wave pairing state
with odd frequency parity\cite{odd-s-wave} can not be ruled out.
 
In using QMC simulations to find superconducting instabilities via
the pair correlation function approach 
one thus faces at least three major difficulties:
(i) long-range pair correlations develop only at low 
temperatures, of order $T_{\rm c}$; 
(ii) it requires large lattice sizes to detect such
long-range correlations in finite systems 
(a problem compounded at finite $T$ in two dimensions by 
the expected algebraic decay of the correlations); and,
(iii) depending on the structure of the superconducting order
parameter, the simulated pair correlation function may provide 
only an undetectably weak signal if one chooses the 
``wrong'' pair creation operators (and one doesn't know
{\it a priori} which operators are the ``right'' ones).
A potentially promising route to circumvent some of these
difficulties is to focus instead on the underlying 
exact effective pairing {\it potential},
{\it i.e.} in precise diagrammatic terms, on the irreducible
particle-particle vertex function.\cite{bulut1,bulut2,scalapino-tjp,bulut0}
The basic idea here is that, in contrast to long-range 
pair correlations, the effective pairing potential 
may be (i) of short range and may (ii) develop strong
signatures of the pairing attraction already at high temperatures.
Also, (iii) by solving for the dominant pairing eigenvalues and eigenfunctions
of the the corresponding particle-particle ladder equation,
the particle-particle vertex approach automatically reveals the symmetry
and space-time structure of the dominant superconducting order parameter.

The existing QMC results for the exact irreducible particle-particle vertex
of the 2D Hubbard model near $1\over2$-filling suggest that
there is indeed a noticeable 1st neighbor pairing attraction and
that a pairing eigenfunction of spin-singlet
$d_{x^2-y^2}$ symmetry becomes dominant
({\it i.e.}, develops the largest pairing eigenvalue) as the temperature
is lowered.\cite{bulut1,bulut0}
In addition, sub-dominant pairing eigenvalues of comparable magnitude 
with eigenfunctions of
odd-frequency triplet $s$- and 
odd-frequency singlet $p$-wave symmetry are found.\cite{bulut1}
The effective ``on-site'' interaction extracted from the irreducible
particle-particle vertex is found to be repulsive.\cite{bulut2}

We should emphasize here that, aside from some preliminary
studies at $1\over2$-filling,\cite{bulut0}
all of the foregoing exact irreducible 
particle-particle vertex results were
obtained at a doping concentration of 
$x\cong 13\%$ and, at that concentration, they are {\it not} 
inconsistent with our results for the pairing $\lambda$-parameters 
shown in Fig.~\ref{fig:v_s-x}(c). 
At $x=13\%$, our $d_{x^2-y^2}$-$\lambda$
is indeed the dominantly attractive one; the $s$-wave-$\lambda$ is 
repulsive at $x=13\%$ and becomes attractive only for $x>15\%$.

Also, unfortunately, the reported results for 
the real-space interaction strengths extracted 
from the irreducible particle-particle vertex \cite{bulut2}
were reported for smaller Hubbard $U$, $U=4t$, whereas 
a physically more realistic larger value of $U=8t$
was used in our calculations of $V_{\rm S}(r,i\omega)$. 
Recall here that, as discussed above,
the on-site overscreening is a large-$U$ effect and
$U=4t$ may simply be too small to produce on-site
overscreening. Interestingly, from a comparison of
the pairing eigenvalues extracted from the 
irreducible particle-particle vertex at $U=4t$ and at 
$U=8t$,\cite{bulut1} it  appears that the (odd frequency tripet)
$s$-wave eigenvalues are {\it increasing} with $U$. 
This observation is not inconsistent with the notion that 
one moves towards on-site overscreening with increasing $U$.

It would therefore be of considerable interest to check whether the
irreducible particle-particle vertex develops an on-site
attraction and stronger $s$-wave pairing tendencies at larger doping,
$x>15\%$, {\it and} larger Hubbard repulsion $U$, {\it e.g.}, for $U=8t$.
This would provide a very stringent test as to whether our
screened potential $V_{\rm S}$ in the charge representation
indeed provides a reasonable approximation
to the exact irreducible particle-particle vertex or whether
particle-particle vertex corrections of higher order in $V_{\rm S}$
are important.

In analyzing the irreducible particle-particle vertex,
it was also found that its momentum and real-space structure,
and the $T$-dependence of that momentum and real-space structure, 
closely resembles that of an appropriately
defined spin fluctuation mediated pairing potential
of the Hubbard model, suggesting that spin
fluctuatons are indeed responsible for the $d$-wave
pairing attraction.\cite{bulut2,scalapino-tjp}
It is possible that the positive near-neighbor charge correlations 
[{\it i.e.}  $\chi(r,i\omega=0)>0$ at near-neighbor $r$-vectors], 
responsible for the $d$-wave pairing attraction in our approach, 
are closely (and perhaps causally) related to short-range 
antiferromagnetic spin correlations 
in Hubbard systems near $1\over2$-filling.
The charge representation approach developed here 
may thus provide a description of the physics
in near-${1\over2}$-filled Hubbard systems
which is complementary to that of a spin 
fluctuation-based approach.\cite{scalapino,bickers,esb}
The overscreening of the on-site potential, 
and hence the possibility of $s$-wave pairing 
in the Hubbard model, is one aspect of this problem 
which may be (with all the above-stated caveats !) ``obvious'' 
in the former, but difficult to reproduce in the latter approach. 
The relationship between our charge representation
formulation and the spin fluctuation picture of
the Hubbard model \cite{scalapino,bickers,esb}
needs to be further investigated.
\section{Summary}
\label{sec:summary}
In summary, by a combination of diagrammatic
and quantum Monte Carlo techniques 
and by exact analytical approaches,
we have studied the dielectric screening and the screening of
electron-electron Coulomb repulsions in the charge 
representation of the 2D Hubbard and of the quasi-2D extended
Hubbard models relevant to the cuprate superconductors.

Applying a combination of QMC simulations and
diagrammatic techniques, we have calculated
the contribution of the quasi-2D extended 
Hubbard electron system at $1\over2$-filling to the 
long-wavelength in-plane dielectric screening in the
undoped insulating cuprate parent compounds.
Combining this result with experimental data
for the observed long-wavelength dielectric tensor
in the $La_2CuO_4$, we have then obtained an 
estimate for the strength of the dielectric 
screening due the insulating ``background'' 
of non-Hubbard electrons and phonons and, thereby,
for the effective strength of the extended $1/|r|$
Coulomb interaction potential in the Hubbard conduction
electron system. Our results imply that no more
than $25\%$ of the observed electronic 
in-plane dieclectric constant ($\epsilon_\infty$)
arises from the $1\over2$-filled 
Hubbard conduction electron system. 

Using again a combination of 
diagrammatic and QMC techniques, 
we have then studied the effective, screened
electron-electron interaction potential
$V_{\rm S}$, as a function of doping $x$,
in the both the 2D Hubbard and quasi 2D extended
Hubbard model.
We find that finite doping gives rise to
overscreening effects which cause $V_{\rm S}$ to
become attractive at finite doping concentrations.
Specifically, we find that the  low-frequency
extended (1st, 2nd... neighbor) part of 
$V_{\rm S}$, while repulsive and 
rapidly suppressed with doping at small $x$,
becomes attractive for $x\gtrsim 5\%$.
Near $x\!\sim\!10-15\%$, the extended part of $V_{\rm S}$ 
exhibits maximum attraction strength and, for $x\gtrsim20-25\%$,
it becomes repulsive again and increases with $x$.
At larger doping, $x\!\gtrsim\!15\%$, 
even the on-site part of $V_{\rm S}$ changes sign
and becomes attractive, with a monotonic
$x$-dependence between $x=0$ and $x=30\%$.
 
Both spatially extended (1st, 2nd, ... neighbor)
and on-site overscreening effects are robust against 
the presence of and moderate changes in the
3D extended $1/|r|$ Coulomb repulsions,
and largely independent of the arrangement
of $CuO_2$-layers in the 3D crystal structure.
Generally, at doping concentrations larger than $5-10\%$,
our results for the screened interaction
potential of the extended Hubbard model, with
extended $1/|r|$ Coulomb interactions of realistic strength,
differ from those of the pure 2D Hubbard model 
(with only on-site $U$ repulsion) by
no more than $10-20\%$. This suggests that the extended
Coulomb effects do not qualitatively alter the
the short-range charge correlations and screening
in the Hubbard electron system at finite doping near 
$1\over2$-filling. In particular the overscreening 
effects which we find at doping concentrations 
$x\gtrsim5\%$ are intrisic to the pure 2D Hubbard model.
These overscreening effects are modified, but neither
caused nor destroyed by the extended Coulomb terms.
When included exactly in the screening calculation,
a repulsive 1st neighbor extended Coulomb term enhances
the 1st neighbor overscreening in $V_S$ at finite doping,
suggesting the possibility that repulsive extended 
Coulomb terms could actually contribute to an extended
pairing attraction.
 
Independent of the QMC results, we have presented an
exact analytical proof that, in the large-$U$ limit, 
doping induced charge fluctuations must give rise to an 
``overscreening'' effect which causes the {\it exact} 
screened on-site Coulomb potential $V_{\rm S}$ in the charge
representation to become attractive.
We have shown that this is indeed an intrinsic 
local property of both large-$U$ Hubbard and extended 
Hubbard models at finite doping which exists independent
of dimensionality, 3D crystal structure or system size.
Our QMC results thus show that this asymptotic large-$U$
scenario is indeed realized in the finite-$U$
parameter range relevant to the cuprates.
We have also shown that on-site overscreening implies
singularities in the imaginary-frequency dependence
of the irreducible polarization insertion and of
its underlying 3-point vertex function.

When analyzed as an effective pairing potential
{\it i.e.} formally, as a 1st order approximation
to the irreducible particle-particle vertex,
the screened potential $V_{\rm S}$ gives rise to a 1st neighbor
$d_{x^2-y^2}$ pairing attraction in the doping range
between $x\gtrsim5\%$ and $x\lesssim20-25\%$
and to an on-site $s$-wave pairing attraction 
for $x>15\%$. The doping dependence of the $d_{x^2-y^2}$
Eliashberg $\lambda$-parameter closely tracks that of the
extended (1st neighbor) attraction of $V_{\rm S}$, with
a maximum of pairing strength near $x=10-15\%$,  
reminiscent of the observed doping dependence of 
the superconducting $T_{\rm c}$ in the cuprates.
The doping dependence of the on-site $s$-wave 
$\lambda$-parameter closely tracks that of the
on-site $V_{\rm S}$, increasing monotonically with $x$
and becoming strongly attractive for $x\sim15\%$.
This suggests the possibility of a doping induced 
transition or cross-over from $d$- to $s$-wave pairing.
We have argued that retardation plays an essential role
in the on-site overscreening of the interaction potential
and that this should be reflected in the temporal /
frequency structure of the corresponding $s$-wave
superconducting order parameter.
\acknowledgments

We would like to acknowledge discussions
with D. Emin, G. Esirgen, K. Levin, 
D. Scalapino, S.K. Sinha, and M.G. Zacher.
This work was supported at the University of Georgia
by NSF Grants No. DMR-9215123 and DMR-9970291,
in part, at the University of California at Santa Barbara
by NSF Grant No. PHY99-07949, and at Universit\"at W\"urzburg
by BMBF (05SB8WWA1) and by DFN Contract No. TK 598-VA/D03.
Computing resources from UCNS at the University of Georgia,
from HLRZ J\"ulich and from HLRS Stuttgart are gratefully acknowledged.

\begin{references}
\bibitem{ext-s-wave}
The non-interacting Fermi surface of
the in-plane 1st neighbor tight-binding band 
is a locus of $k$-points where $\cos(ak_x)+\cos(ak_y)={\rm const}$.
The in-plane 1st neighbor extended $s$-wave pair 
wavefunction $\eta_{s1}(k)=\cos(ak_x)+\cos(ak_y)$
is therefore constant on that Fermi surface and 
gives exactly the same value for the EM $\lambda$-parameter 
as the on-site $s$-wave pair wavefunction $\eta_s(k)=1$.
\end{references}
\end{document}